\documentclass[preprint2]{aastex}

\slugcomment{attempt at mexico proceedings w/o badness 2006 apr 20}

\shorttitle{Jet Constraints}
\shortauthors{Harris \& Krawczynski}

\begin{document}


\title{Constraints on the Nature of Jets from kpc Scale X-ray
Data\footnote{This paper is based on a poster contribution to the
meeting, ``Triggering Relativistic Jets'', held in Cozumel, MX at the
end of March 2005 and will be published via a CD distributed with a
special issue of Revista Mexicana de Astronomia y Astrofisica, Serie
de Conferencias, eds. W.H. Lee \& E. Ramirez-Ruiz, 2006.}}

\author{D. E. Harris}

\affil{Smithsonian Astrophysical Observatory, Cambridge MA USA}

\and

\author{H. Krawczynski}

\affil{Washington University in St. Louis, MO, USA}


\begin{abstract}
Motivated by the large number of jets detected by the Chandra X-ray
Observatory, and by the inverse Compton X-ray emission model (IC/CMB)
for relativistic jets, we revisit two basic questions: ``If the medium
that carries the jet's energy consists of hot electrons, can we use
the physical length of the jet to constrain the maximum electron
energy?'' and ``Why do jets have knots?''  Based on the two
non-thermal emission processes for X-rays from jets, we consider
constraints on the jet medium and other properties from these two simple
questions.  We argue that hot pairs cannot be the dominant constituent
of the medium responsible for the jet's momentum flux and that some
mechanisms for producing fluctuating brightness along jets (rather
than a monotonically decreasing intensity) are precluded by observed
jet morphologies.
\end{abstract}






\section{Introduction \label{sec:intro}}  

The impetus for this contribution arises from the uncertainty as to
the X-ray emission process from kpc scale jets for powerful (FRII)
radio galaxies and quasars.  Although the current consensus is that
FRI radio jet emission is dominated by the synchrotron process from
the radio to X-ray frequencies, most papers dealing with quasar jets
ascribe the X-ray emission to inverse Compton emission from the normal
power law (or broken power law) distribution of relativistic electrons
responsible for the radio and optical synchrotron emissions,
scattering off photons of the cosmic microwave background (IC/CMB).
This model relies on the bulk velocity of the jet medium having values
close to the speed of light, so that the effective energy density of
the CMB is augmented by the square of the jet's Lorentz factor,
$\Gamma$ (Celotti et al. 2001, Tavecchio et al. 2000, Harris \&
Krawczynski 2002, Sambruna et al. 2002, 2004).  Typical values of
$\Gamma$ quoted in the literature lie in the range 5 to 30.

There are several notable problems for the IC/CMB model (Stawarz 2004,
Dermer \& Atoyan 2004), so we also consider constraints derived from
values of $\Gamma$ expected for synchrotron models (i.e. $\Gamma$ of
order 3 to 5 instead of 10 or greater).

A separate, but related problem is the mechanism that produces
brightness changes along the jet, i.e. the structures we normally call
'knots'.  We will discuss several mechanisms that might be
responsible for knots in light of the radio/X-ray morphologies of jets.

\section{If the medium that carries the jet's energy consists of hot
  electrons, can we use the physical length of the jet to constrain
the maximum electron energy?}

Disregarding the energy of the electrons producing the observed
emission, we consider what the 'medium' might be that is responsible
for transporting the energy of the jet:

\begin{itemize}

\item{a normal proton/electron plasma}


\item{Poynting Flux}


\item{a pair dominated plasma}


\end{itemize}

Regardless of the magnetic field strength, any 'hot'
electrons will suffer inescapable inverse Compton losses to the 
photons of the microwave background (extremely energetic electrons for
which IC losses are suppressed by the Klein-Nishina cross section are
precluded by even extremely weak magnetic fields).  Simply by observing
emission at the end of jets, we can calculate the 'age of the medium',
i.e. how long the various E$^2$ energy losses have been operating.  In
this way, we can find the maximum permissible Lorentz factor,
$max(\gamma$), for the pair dominated case.

\begin{figure}[!t]   
\includegraphics[width=\columnwidth]{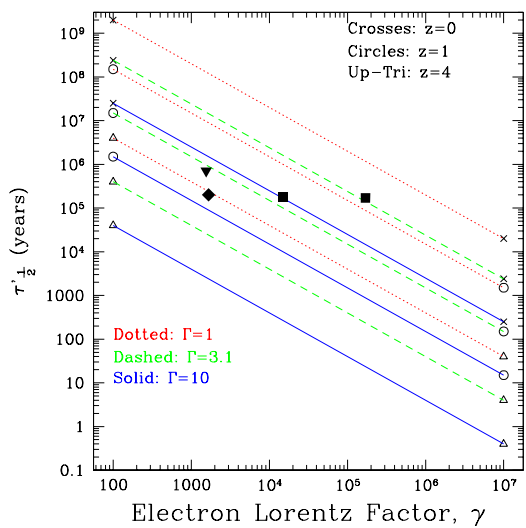}
\caption{The maximum half-life for relativistic electrons determined
  by the mandatory energy losses from inverse Compton scattering off
  photons of the CMB.  
  Dotted lines are for jets which are not moving
  relativistically and are shown for reference.  Dashed lines are for
  mildly relativistic bulk velocities ($\Gamma$=3.1) and solid lines
  are for $\Gamma$=10.  The half life is given in the jet frame.
  Three characteristic values of the redshift are given for each
  $\Gamma$.  The ages of the jet medium at the ends of 3 jets are also
  plotted.  For 3C273 (solid squares) we give two values: the one to
  the right corresponds to $\Gamma$=3, $\theta$=20$^{\circ}$ while
  that to the left is for $\Gamma$=10, $\theta$=5$^{\circ}$.  For
  PKS0637 (diamond) we assumed $\Gamma$=10, $\theta$=5$^{\circ}$.  The
  down triangle indicates the age for the medium at the end of the jet
  of PKS1127 with $\Gamma$=3, $\theta$=20$^{\circ}$.}
\label{fig:half}
\end{figure}

The 'Half-life' plot shown (fig.~\ref{fig:half}) is essentially 9
versions of eq.(B5) of Harris \& Krawczynski (2002).  A simplified
version of this equation for the half-life of electrons in the jet
frame is:

\smallskip

\centerline{$\tau'=\frac{10^{13}}{\gamma'[B'^2 +
      40\times\Gamma^2\times(1+z)^4]}$ (years)}

where B' is in $\mu$G.

 We take 3 values of the bulk Lorentz factor for the jet: $\Gamma$=1
(no beaming, just for reference), $\Gamma$=3.16 (a typical value for
synchrotron models), and $\Gamma$=10 (the classic solution for the
PKS0637 IC/CMB model).  For each of these we show 3 characteristic
values of the redshift.  Since we were interested in the largest
possible value of $\tau$, we took only the CMB energy density
and set the magnetic field strength to 3 $\mu$G.  In reality, B' will
most likely be significantly larger than this value over at least
parts of the jet, and IC losses will be more severe than indicated for
the initial parts of the jet where starlight and/or quasar radiation
probably exceeds the CMB in energy density.

To calculate how old the jet medium is by the time it reaches
the end of the jet.  We take the projected length, divide that by the
most likely value of sin~$\theta$ ($\theta$ is the angle between the
l.o.s. and the jet axis); convert to light years; and divide by
$\Gamma$.  With this age for the jet medium (in the jet frame), we know
that any surviving electrons must have $\gamma$ less than the value
corresponding to the halflife calculated for that particular jet
(i.e. the appropriate values of z and $\Gamma$).  

For synchrotron models we take characteristic values of $\Gamma$=3,
$\theta$=20 (typical parameters which can hide the counterjet;
e.g. M87, see Harris et al. 2003) and for the IC/CMB model we take
larger values of $\Gamma$ and smaller $\theta$.  We show 3 examples:
3C273, PKS0637, and PKS1127.

For 3C273, we take the most likely values for IC/CMB of $\Gamma$=10
and $\theta=5^\circ$.  These conditions yield a $max(\gamma$) of
15,000.  For synchrotron models with relaxed beaming conditions,
$max(\gamma)\approx~2\times10^5$.  These two values are shown in
fig.~\ref{fig:half}.

In the case of PKS0637, stronger limits could be found for the end of
the radio jet, but we use the distance of the strong radio/X-ray knots
8$''$ from the quasar.  With $\Gamma$=10 and $\theta=5^\circ$, we find
a $max(\gamma$) value of 1700.

PKS1127 has a redshift of z=1.16 so beaming models do not require a
large $\Gamma$ (Harris \& Krawczynski 2002, Siemiginowska et al. 2002).
Knot C is located 28$''$ from the core.  For this source, there is not
much difference between synchrotron and IC/CMB models insofar as our
analysis is concerned.  For $\theta=20^\circ$ and $\Gamma$=3,
$max(\gamma$) is 1600.

These limits on $\gamma$ are sufficient to convince us that 'hot'
pairs are not a viable candidate for the agent responsible for the
energy/momentum flow of powerful jets.  Since we find similar
constraints for PKS0637 and for PKS1127, this conclusion does not rely
on models that require large values of $\Gamma$.

\section{Why do jets have knots?}

\subsection{Synchrotron Models}

In this section we will consider knots in both low power and high
power jets.  Conventional wisdom has it that knots [a.k.a. marked
brightness enhancements] occur because internal shocks accelerate
particles, and these particles radiate.  Good examples are M87/knot A
and 3C120/k25 which show sharp gradients in radio brightness, often as
an inclined linear feature.

\begin{figure}[!t]    
\includegraphics[width=\columnwidth]{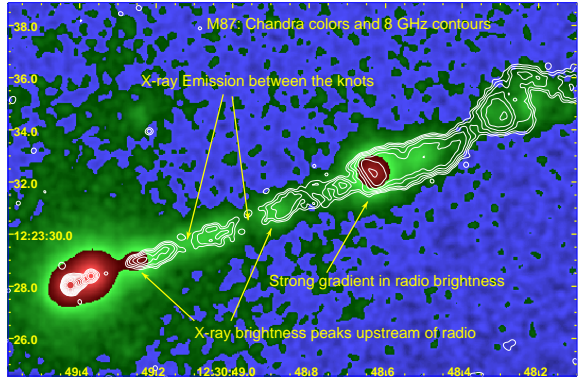}
\caption{A Chandra image of the M87 jet, with radio contours
  overlayed.  The effective resolution of the X-ray data is about
  0.6$''$ FWHM, whilst that of the radio is 0.24$''$ FWHM.  With
  matching beams, the features illustrated do not
  change.\label{fig:m87}}
\end{figure}

However, there is also X-ray emission between the radio knots
indicating that there must additionally be some distributed
acceleration process to generate electrons with $\gamma\approx~10^7$
wherever X-ray emission is found (see fig.~\ref{fig:m87}).  This
follows from the very short half-life (of order a year) of the
electrons responsible for synchrotron X-rays.

\subsection{IC/CMB with beaming}

The main question for the IC/CMB model is why don't X-ray 'knots',
once they appear, trail off downstream more gradually than the radio
and optical since for IC/CMB, the half-life for the X-ray emitting
electrons ($\gamma\approx$100) is very much longer than for those
producing optical and radio emission.

\subsection{General processes for producing knots}

\begin{figure}[!t]   
  \includegraphics[width=\columnwidth]{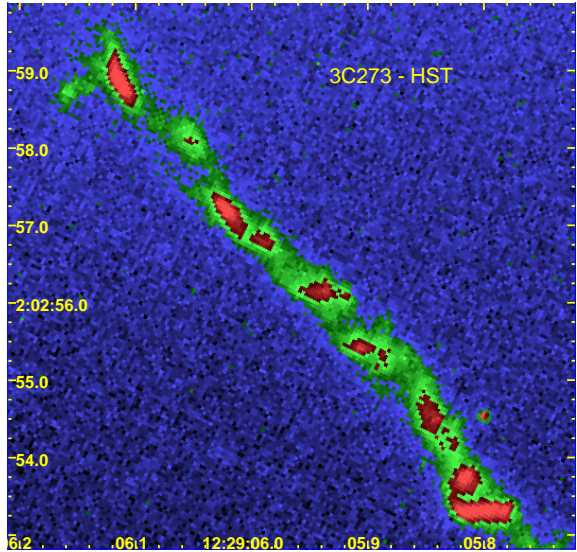}
\caption{An HST image of 3C273\label{fig:273}}
\end{figure}

\begin{itemize}

\item{Doppler boosting: if the jet medium follows a curved trajectory,
  (e.g. a helix as proposed for VLBI scale jets by Gabuzda, Murray, \&
  Cronin (2004), Asada et al. (2002) and Hong et al. (2004)), we might
  see only segments of the trajectory for which the angle to the
  l.o.s. is small.  The HST image of 3C273 (the kpc scale jet is shown
  in fig.~\ref{fig:273}), resembles the projection of a helix.  This
  would work for either X-ray emission model although the large
  $\Gamma$'s required for IC/CMB would mean that these jets would have
  higher contrast than lower $\Gamma$ (synchrotron) jets like M87.}

\item{Intermittent Ejection from the central engine - which would mean
that kpc scale knots are moving, like pc scale blobs.  This also works
for both emission models.}

\item{Acceleration and Deceleration - changes $\Gamma$ so that more
  or less IC X-rays are produced because the effective photon energy
  density goes as $\Gamma^2$.  This process would operate only for the
  IC/CMB model, but is most likely not feasible because any
  significant increase in $\Gamma$ would require a large energy
  source.  Furthermore, at the location of internal shocks where the
  radio emission is high (e.g. the radio knot A in the M87 jet) we
  would expect a deceleration of the jet medium leading to less X-ray
  IC emission, contrary to the observed bright increase in X-ray
  emission.}

\item{Massive expansion/contraction - If the disappearance of a knot
  is to be explained by expansion (which would certainly lower the
  emissivity for both models), we would expect a marked change in the
  ratio of IC to synchrotron emission.  This follows because although
  the electron energy distribution, N(E), will suffer a uniform drop,
  there
  will also be a very strong effect of lowering the magnetic field
  strength: the synchrotron emissivity will decrease as $B^2$ and a
  fixed reception band will be sampling a higher energy segment of the
  N(E) power law which will have a smaller amplitude.  Thus we would
  expect a sharper decrease of the synchrotron emissivity (radio and
  optical) than the IC emissivity (X-ray).  Just the opposite is
  actually observed in many cases.}

\end{itemize}

\section{Summary}

In both the synchrotron and IC/CMB emission models, hot electrons
cannot be the main carrier of jet energy and momentum.  That leaves
Poynting flux, 'cold' electrons/positrons, or protons (hot or cold).

In the table below we summarize the situation for generation of
knots.  If the IC/CMB process were responsible for X-ray emission from
powerful jets, then the most favored knot processes would be curved
trajectories and/or intermittent ejection.  If the X-rays come from
synchrotron emission, then two additional processes are viable:
internal shocks and expansion/contraction.

\begin{center}
\footnotesize
\begin{tabular}{lccc}

\hline\hline
\multicolumn{1}{c}{Mechanism} & \multicolumn{1}{c}{Sync.}      &
\multicolumn{1}{c}{
Key Element}      &  \multicolumn{1}{c}{IC/CMB    } \\
\hline\hline
internal shocks    &   Y    &  offsets     &  N     \\
Distrib. Accel.    &   Y    &   X-ray emis.    &  not required     \\
 Curved Traject.   & Y      & contrast      &  Y(?)     \\
Intermittent Eject.    &   Y    &  (vlbi blobs)     &   Y    \\
Accel./Decel.    &   N    &  source of energy     &   N(?)    \\
Expand/Contract    &  Y     & offsets      &   N    \\

\end{tabular}

\end{center}

\normalsize

The classical explanation of knots as internal shocks does not account
for the brightness differences between radio, optical, and X-ray
images under the IC/CMB model, but is fully consistent with the
synchrotron model.  The only two knot production methods which we find
to be consistent with both X-ray emission models are the intermittent
ejection and curved trajectory scenarios (these are not mutually
exclusive).

\acknowledgments 

We thank C. Cheung for helpful comments on the manuscript.  A list of
jets detected in the X-ray band is available at:
\url{http://hea-www.harvard.edu/XJET/}.  This work was partially
supported by NASA contract NAS8-39073 and grants GO2-3144X, GO3-4124A,
and GO4-5131X.  HK acknowledges support by the DoE in the framework of
the Outstanding Junior Investigator program.

\end{document}